# Parity-Time Symmetry Breaking in Coupled Nanobeam Cavities

Senlin Zhang[1], Zhengdong Yong[1], Yuguang Zhang[1] and Sailing He[1, 2*]

1 Centre for Optical and Electromagnetic Research, Zhejiang Provincial Key Laboratory for Sensing Technologies, JORCEP, East Building #5,Zijingang Campus, Zhejiang University, Hangzhou 310058, China

2 Department of Electromagnetic Engineering, School of Electrical Engineering, Royal Institute of Technology (KTH), S-100 44 Stockholm, Sweden

* Corresponding author: sailing@kth.se

**Abstract**

The parity-time symmetry (PT symmetry) breaking phenomenon is investigated in a coupled nanobeam cavity system. An exceptional point is observed during the tuning of the relation of the gain/loss and coupling strength of the closely placed nanobeam pairs. The PT symmetry concept can be applied to realize unidirectional light propagation and single mode operation lasers, which may allow for a new way to harness the optical signal in photonic integrated circuits. Otherwise, operating at this particular exceptional point, sensitivity of tiny perturbation detection can be enhanced greatly compared with conventional sensors.

## Introduction

Since the notion that non-Hermitian Hamiltonians respecting PT symmetry can exhibit entirely real spectra was proposed by Bender et al. in 1998 [1], the PT symmetry concept has been extensively researched. Generally, a Hamiltonian is PT symmetric provided that it commutes with the PT operator [1], that is, its complex potential is subject to a spatial symmetry constraint $V(x) = V^*(-x)$. This complex potential can be easily implemented in the optical realm by modulating the real part and imaginary part of the refractive index to be an even function and odd function, respectively. For example, in the proposed coupled nanobeam cavity system (Fig. 2(b)), the real part and imaginary part of the refractive index can be artificially tuned to be even-symmetric and odd-symmetric along the vertical direction, respectively. Since the imaginary part of the refractive index represents gain or loss, it is easy to realize the above constraints through tuning the gain/loss level of an optical system.

One of the most fascinating features of a PT symmetric system is the existence of an exceptional point (EP), above which the PT symmetry is broken and the real spectra start to become complex. The appealing optical phenomena has resulted in much investigation and implementation in optics, both theoretically and experimentally. Basic theory and analytic results of PT symmetric structures are studied successively [2-3]. Optical systems including coupled optical waveguides [4-6], photonic lattices [7-10], plasmonics [11], pumped lasers at EP [12-15], nonlinear PT symmetric systems [16-17], chaotic optical microcavity [18], atom cavity composite system [19], and whispering gallery modes [20-21] have been proposed and explored. Applications such as laser absorbers [12-15], single mode lasers [22-23], unidirectional light propagation [24-26], and sensitivity enhanced particle detectors [27] have been validated.

Despite the extensive research, the PT symmetry phenomenon has not been discussed in nanobeam cavities, which have high quality factors (Q factors), small mode volume, compatibility with standard complementary metal-oxide semiconductor (CMOS) technology and ease of coupling with photonic integrated waveguides [28].

In this paper, we propose a coupled nanobeam cavity system that shows a sharp symmetry transition. The system is based on a silicon-on-insulator (SOI) platform and the nanobeam cavity is designed through the deterministic design procedure proposed by Quan et al. [29]. The factors determining the PT symmetry transition are considered, and potential applications such as optical isolators, lasers and sensing are also discussed.

## Structure of nanobeam cavity

The parameters of a one-dimensional photonic crystal nanobeam cavity (PCNC) are determined according to Ref. [29]. The thickness and width of the silicon core are 220nm and 700nm, respectively. The refractive index of the silicon core is 3.47 while that of the silica buffer layer and the polymer cladding are both 1.46. There are a total of 38 holes etched into the silicon layer with 10 holes whose dimensions linearly decrease from the center to both ends and 9 following holes with constant radii. The lattice constant of the PCNC is 330nm. The geometry and vital characteristics of the PCNC are illustrated in Fig. 1.

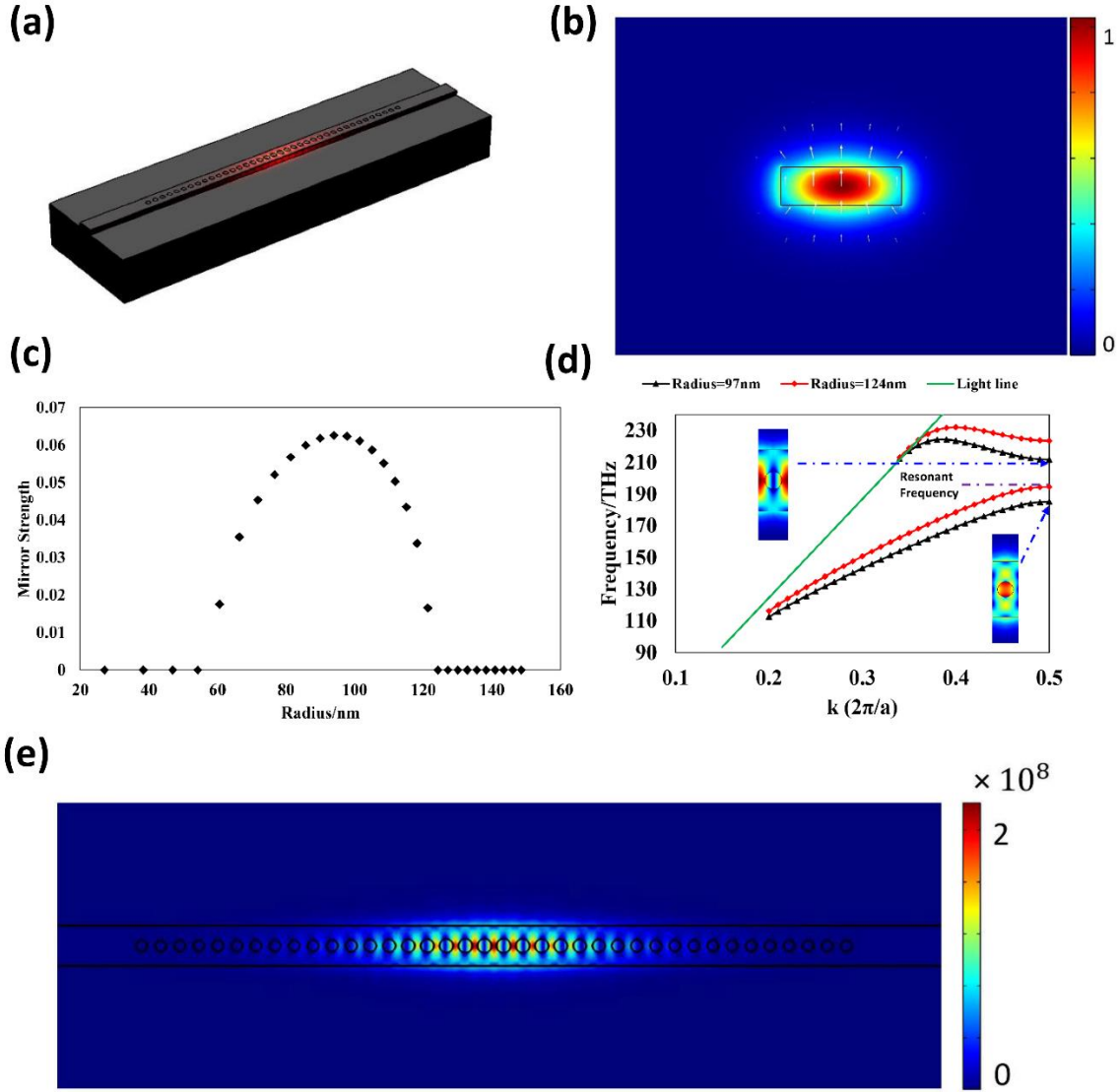

Fig. 1. (a) Geometric structure of the PCNC. The upper cladding is omitted for illustration. (b) The distribution of the electric field of the 220 × 700nm waveguide. (c) Mirror strength versus different radii of holes etched into the silicon core. (d) Bandgaps for radii of 97nm and 124nm of the unit cell. The resonant mode is around the lower band of the 124nm-radius cell. The electric field distribution for the so-called dielectric mode and air mode are shown in the inset. (e) The light intensity of the resonant mode at 190.31THz with a quality factor (Q factor) of about 64k.

The geometry of the device is shown in Fig. 1(a). Fig. 1(b) depicts the fundamental mode of the core waveguide (220×700nm). High Q factor can be achieved if the electric field attenuation shows a Gaussian shape from the center to the outermost part of the cavity along the horizontal direction, where the attenuation coefficient is defined as the so-called mirror strength [29]. In order to determine the size of the holes, series of band diagram simulations were implemented using the Finite Element Method (FEM) to calculate the mirror strength of different radii. The analytic equation defining the mirror strength is $\sqrt{(\omega_2 - \omega_1)^2 / (\omega_2 + \omega_1)^2 - (\omega_{res} - \omega_0)^2 / \omega_0^2}$, where $\omega_1$, $\omega_2$, $\omega_0$ and $\omega_{res}$ are the dielectric band edge, air band edge, middle frequency and resonant frequency [29]. The calculated mirror strength is shown in Fig. 1(c) as the hole radius varies, and the radii of the holes are chosen to be 124nm in the center and 97nm (with the largest reflecting capability) in the end according to Fig. 1(c). The band diagram for the central and outermost holes are shown in Fig. 1(d), and the optical intensity distribution of the dielectric band edge and air band edge are shown in the inset. Fig. 1(e) characterizes the resonant mode at 190.31THz which displays a Q factor of about 64k. One thing to note is that the Q factor can be further elevated (by up to several millions) through increasing the number of holes. However, since the Q factor is not so crucial in the consideration of PT symmetry, only 38 holes in total are chosen for simplicity of numerical calculations.

## Coupled Nanobeam Cavity Pairs

Two closely placed PCNC pairs will show frequency splitting resulting from the coupling between the cavities. The consequent frequency detuning can be derived as $\Delta\omega = \sqrt{(\omega_1 - \omega_2)^2 + 4g^2}$, where $\omega_1$ and $\omega_2$ are the resonant frequencies of the two PCNCs when the distance between them is large enough and g represents the coupling strength [30]. We calculated the frequency detuning of the PCNC pairs under different gaps, and the results are shown in Fig. (2).

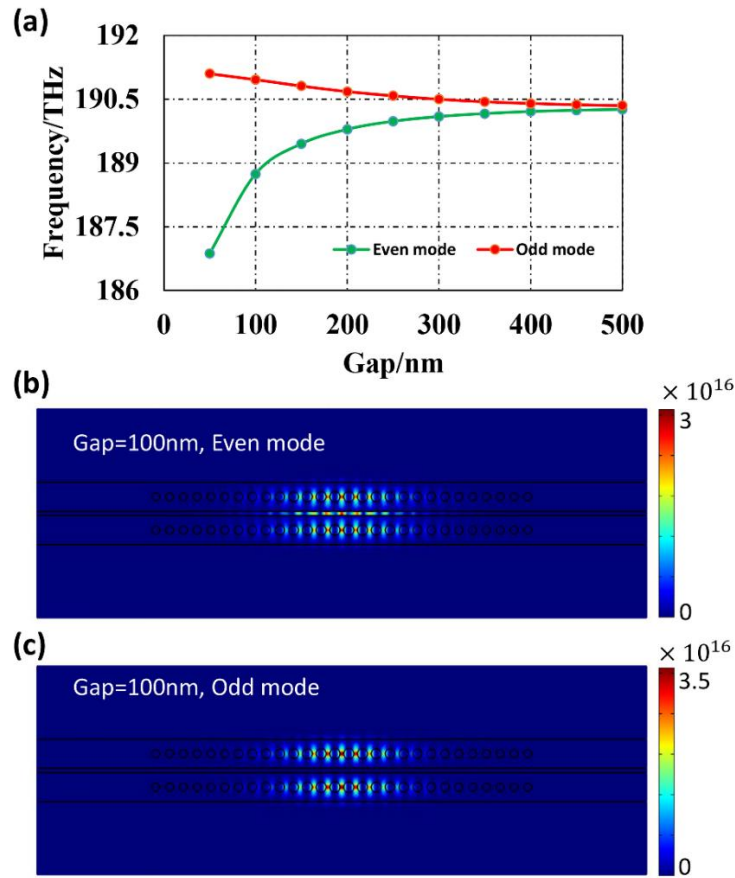

Fig. 2. (a) The frequency detuning with variant gap between the cavities. (b) and (c) depict the even and odd modes, respectively.

Fig. 2(a) manifests the large frequency splitting of the PCNC pairs when placed closely. When the gap is 100nm, the frequency detuning is as high as 2.22THz. As the gap becomes larger, the coupling strength between the cavities declines and leads to smaller frequency detuning. Fig. 2(b) and (c) characterize the two supermodes (the symmetric even mode and antisymmetric odd mode). As we can see from Fig. 2(a), the frequency of the even mode shrinks intensely as the gap increases, while the frequency change of the odd mode is quite mild. Second-order cross- and self-coupling effects account for this phenomenon [31].

## PT symmetry broken in PCNC pairs

For weakly coupled cavities with modal gain/loss, the governing coupled mode equations for the two supermodes are as follows: [2, 30]

$$\begin{cases} \dfrac{da}{dz} = -i\omega_1 a + i\kappa b + g_1 a \\ \dfrac{db}{dz} = -i\omega_2 a + i\kappa a + g_2 a \end{cases} \quad (1)$$

where a and b are the respective resonant modal amplitudes of the two cavities, $\omega_1$ and $\omega_2$ are the related eigenfrequencies, $\kappa$ is the coupling strength between cavities and $g_{(1,2)}$ is the gain/loss of the two nanobeam cavities (positive represents gain and negative for loss). The eigenfrequencies of the system can be derived as: [23]

$$\omega_{(1,2)} = \omega_0 + i\frac{g_1 + g_2}{2} \pm \sqrt{\kappa^2 - (\frac{g_1 - g_2}{2})^2} \quad (2)$$

where $\omega_0$ is the eigenfrequency of the separated nanobeam cavity. Apparently, the eigenfrequencies of the supermodes not only depend on the coupling strength, but also are affected by the gain/loss contrast defined as $(g_1 - g_2)/2$, of the PCNC pairs [23].

Theoretically, the EP is reached when the gain and loss of the coupled cavities system are balanced, and the gain/loss contrast is matched with the coupling strength between the cavities, that is $|\kappa| = |(g_1 - g_2)|/2$. In order to study the PT symmetry broken in the PCNC pairs, the relation between the eigenfrequency of the system and the gain/loss contrast and coupling strength is considered. To ensure that the gain and loss of the system are balanced, the imaginary part of the polymer in one of the PCNC pairs is set to be negative to represent gain while the other is set to a positive value to model loss. The absolute value of the two artificial gain/loss values are set to be equal. Generally, such gain or loss can be achieved through quantum well lasers, photorefractive structures [7] and erbium doping [21, 30] at the C band. The calculated PT symmetry transition under variable gain/loss contrast with specified coupling strength is shown in Fig. 3. Fig. 3(a) and (b) illustrate the evolution of the eigenfrequencies of the two supermodes under variable gain/loss contrast when the gap between the two PCNCs is set to be 200nm. Apparently, the EP is reached when the gain/loss contrast is about 0.1. In such a case, both the real part and imaginary part of the eigenfrequencies coalesce. When the gain/loss contrast crosses the transition threshold slightly, the real part of the eigenfrequencies remains identical while the imaginary part of the eigenfrequencies bifurcate and one of the mode experiences gain (negative imaginary part of eigenfrequency) while the other experiences damping (positive imaginary part of eigenfrequency). Fig. 3(c) shows the optical intensity distribution under the EP conditions. One can see the distinct localization of the supermodes under this condition clearly. This is quite abnormal compared with conventional coupled PCNC pairs. As the gain/loss contrast continues to increase, the "even mode" (amplified mode) experiences a larger amplification while the "odd mode" (lossy mode) keeps damping.

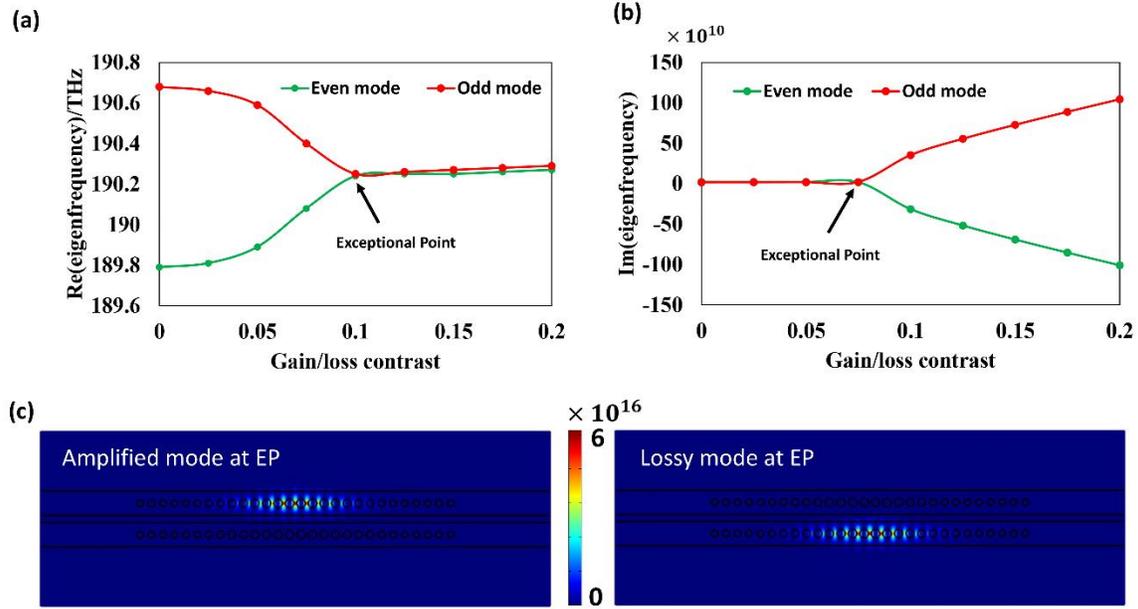

Fig. 3. The calculated PT symmetry transition as the gain/loss contrast is finely tuned. The gain is implanted in the upper PCNC alone while the lower PCNC suffers loss. (a) and (b) depict respectively the evolution of the real and imaginary parts of the eigenfrequencies of the PCNC pairs when gain/loss contrast varies. (The gap between the PCNC pairs is 200nm.) (c) The exceptional supermodes under the so-called EP. The modes experiencing gain and loss are shown in the left and right part of (c) separately.

In order to inspect the relation between the transition point and the coupling strength, the evolution of the eigenfrequencies is also calculated when the gap is 100nm and 50nm. Fig. 4(a) and (b) show the transformation of the real and imaginary part of the eigenfrequencies, respectively, when gap = 100nm while (c) and (d) depict the case when gap = 50nm. A larger transition threshold (gain/loss contrast) is observed as the gap declines.

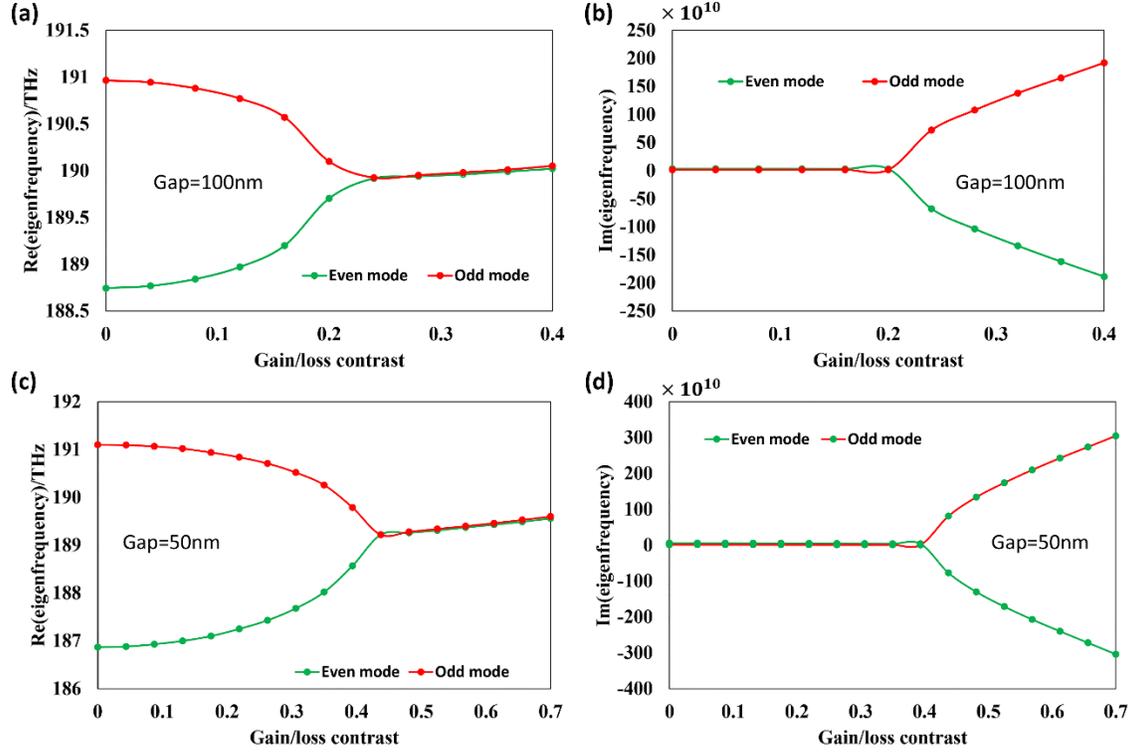

Fig. 4. (a) and (b) illustrate the PT transition when the gain/loss contrast is finely tuned with the gap between the coupled cavities fixed to 100nm, while (c) and (d) are the case when gap = 50nm.

The relation between PT transition and coupling strength is further investigated with a specified gain/loss contrast. A gain/loss contrast of 0.1 is chosen, for we have affirmed that the coupling strength matches this gain/loss contrast when the gap is 200nm. The variable coupling strength is represented by the different gaps between the cavities and numerical calculation results are shown in Fig. 5. From Fig. 5(a), one can see that the exceptional point deviates when the coupling strength is larger than the gain/loss contrast (gap<200nm). Fig. 5(b) shows that a portion of the amplified mode's energy is distributed in the "lossy cavity" and vice versa when the gap is 50nm. On the contrary, when the PT transition threshold (when the coupling strength equals the gain/loss contrast) is crossed, both the amplified mode and lossy mode are well localized in each cavity as shown in the lower two figures of Fig. 5(b) (gap=500nm). The numerical calculations are well in accordance with the theoretical predications [32].

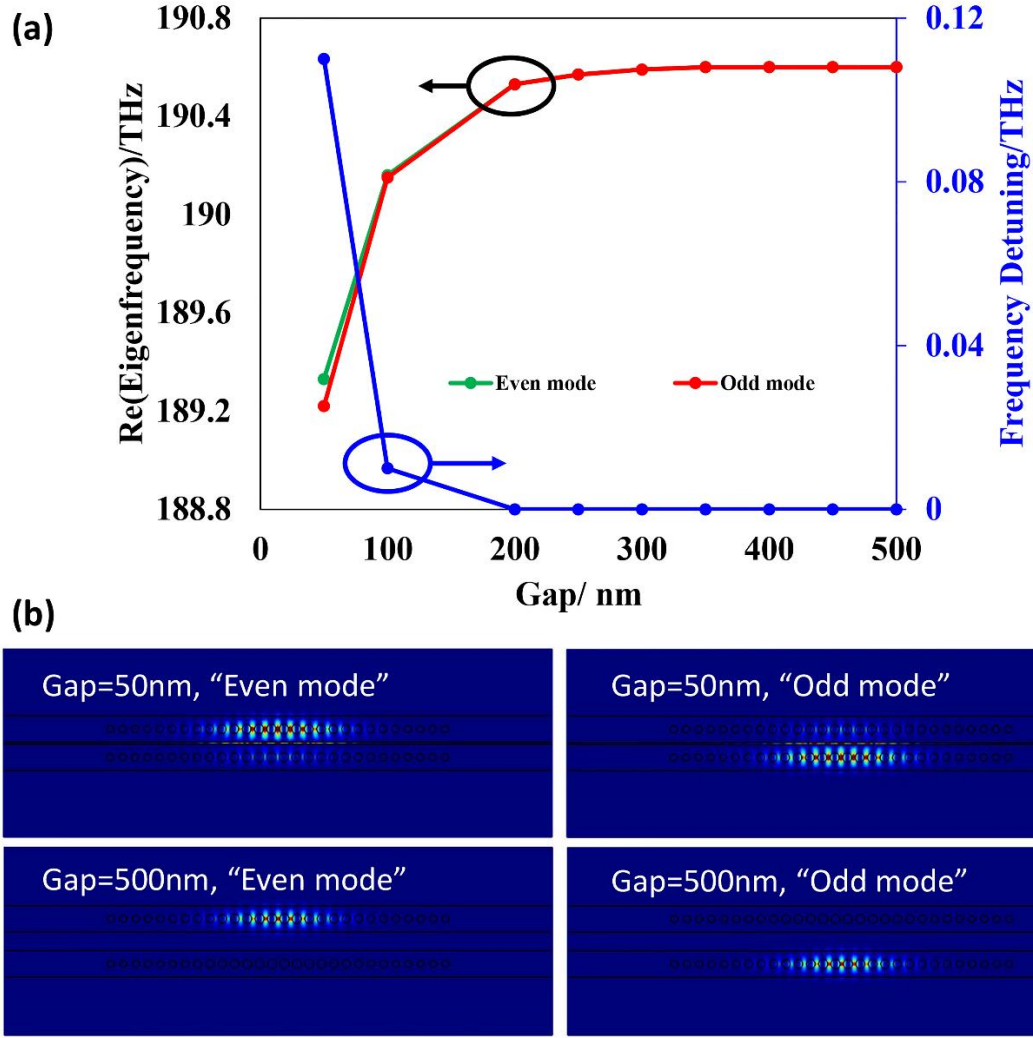

Fig. 5. (a) The relation between PT transition and coupling strength (smaller gap gives larger coupling strength) with balanced gain/loss. (b) Distribution of the electric field intensity under different gaps (50nm and 500nm).

## Unidirectional Light Propagation

Unidirectional light propagation, i.e., optical isolators, has been studied extensively since it is one of the key components for realizing on-chip optical signal processing. The beneficial localization characteristic of the PT symmetric nanobeam cavities makes this a promising concept for achieving unidirectional light propagation.

The numerical results using Finite Difference Time Domain (FDTD) are shown in Fig. 6. Fig. 6(a) depicts the configuration of the system used in the calculation. The direction that light is injected from the left port (port 1) of the passive cavity is defined as the forward direction (1 → 2) while the direction that light is injected from the right port (port 2) of the active cavity is the backward direction (2 → 1). The two eigenfrequencies (fundamental mode) of the cavities used here with no gain are around 1570nm as shown in the inset of Fig. 6(c). When the system is in broken symmetry, the two eigenfrequencies coalesce just as discussed, and the transmission spectrum is shown in Fig. 6(c). Fig. 6(b) and (d) illustrate the unidirectional light propagation feature when the system is in broken-symmetry status. The light is well confined in the active cavity no matter which port (1 or 2) the light is injected into. Thus, when the light is input from port 1 (forward), the output optical intensity received at port 2 is shown in Fig. 6(d). On the contrary, the light that is input from port 2 (backward) is not allowed to transmit to the output port 1. The power detected in port 1when the light is input from port 2 (backward) is quite weak, due to the unbalanced gain/loss contrast. If the gain and loss of the two cavities are judiciously chosen, the backward propagation can theoretically be thoroughly prohibited. The realization of unidirectional light propagation here outperforms previous works [21] in terms of its compact feature size and excellent compatibility with photonic integrated circuits.

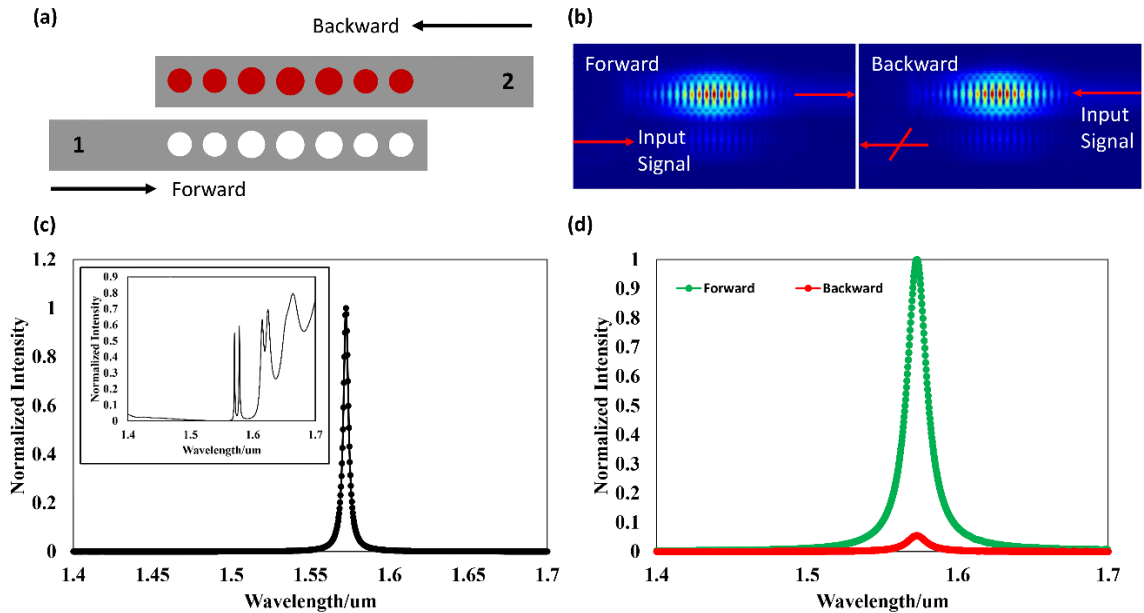

Fig. 6. (a) The configuration of the PT symmetric PCNCs. The upper nanobeam cavity with red holes is the active cavity, and the lower one is the passive cavity. The direction 1 → 2 is defined as the forward direction and the 2 → 1 direction is backward. (b) The electric intensity distribution under the forward and backward light propagation. (c) The transmission spectrum under broken PT-symmetry (forward). The inset shows the transmission spectrum when the two coupled cavities are both passive. (d) The transmission spectrum when the light is injected in the forward (green line) and backward (red line) directions.

Another interesting phenomenon shown in Fig. 6 is the potential application of the system in single mode laser operation. To overcome the nonlinear effect due to the large Purcell Factor of cavities, one method is to increase the feature size of the cavities. However, higher order modes arise as a result, which has detrimental effects on the performance of the lasers [22-23]. The concept of PT symmetry has been exploited to realize single mode operation lasers. Since the higher order modes have stronger coupling strength than the fundamental mode, it is easier to reach the transition threshold for the fundamental mode. In other words, it is possible to make the fundamental mode experience gain while the higher order modes remain neutral. From Fig. 6(c), one can see that besides the coalescence of the two fundamental modes, the higher order modes are totally suppressed in the output spectrum. Theoretically, the higher order transverse modes can also be filtered when the laser is set to work at PT symmetric conditions [23].

## Enhanced sensitivity for single particle induced perturbation detection

Many entirely passive nanobeam cavity have been implemented to sense the perturbation of the environment around the cavity based on the detection of the shift of frequency. However, the sensitivity of these sensors is proportional to the disturbance ($\varepsilon \sim$ the effective refractive index perturbation in most cases). We will show that by operating at the EP the sensitivity can be enhanced massively (sensitivity $\sim \sqrt{\varepsilon}$) due to the sharp transition in the vicinity of the EP [27]. Therefore, the sensors based on the PT symmetry broken concept can be implemented to detect tiny perturbations induced by a single particle and show better performance than conventional sensors.

In order to examine the functionality of the PT-symmetric sensors, the coupled PCNC pairs with a gap of 50nm working at EP are utilized. Since the disturbance has a complex value in practical situations [27, 33], we use a single gold particle as the target for sensing. Larger gold particle will results in larger disturbance (larger loss which will break the balance between the coupling strength and gain/loss contrast more seriously) and leads to larger frequency splittings. Fig. 7(a) shows the performance comparison between PT-symmetric configured sensors and conventional sensors. The advantage of PT-symmetric sensors over conventional sensors can be clearly seen from the drastic change of frequencies due to the variation of the geometric size of the gold particle. The sensitivity of the PT-symmetric sensors is more than twice as large as that of the conventional sensors when the radius of the gold particle changes from 20nm to 80nm. Theoretically, the sensitivity not only depends on the disturbance, but also is affected by the gain/loss contrast ($\sim \sqrt{g\varepsilon}$)) in the PT symmetric sensors [33]. Fig. 7(b) shows the situations when the coupling strength of the PT symmetric sensors varies. The calculated results show that the sensors with a stronger coupling strength, i.e., the system needs a larger gain/loss contrast to realize the broken PT symmetry, possess a higher sensitivity. In addition, the capability of enhancing the sensitivity of PCNCs applies to other judiciously designed ultrahigh-sensitive PCNC sensors provided that the PCNCs are tuned to work at the EP.

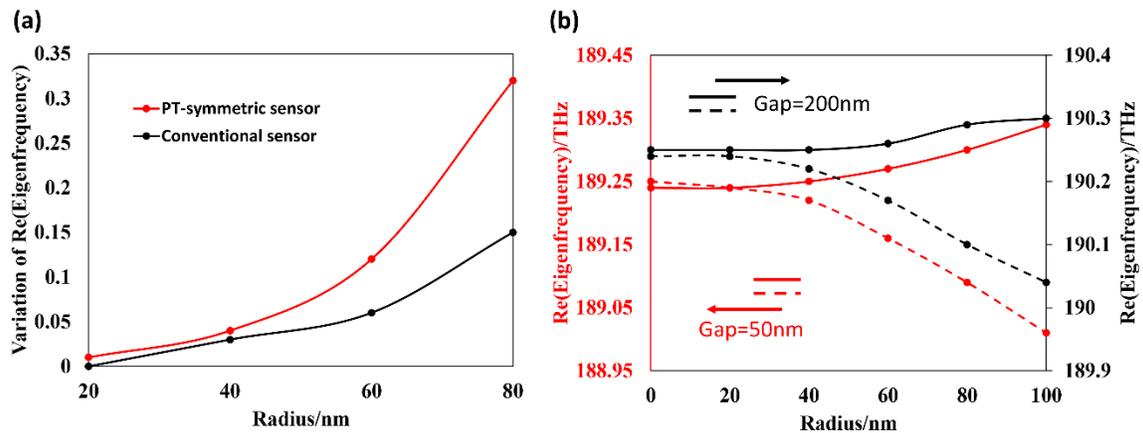

Fig. 7. (a) The drastic frequency change of sensors against single particle perturbation in PT symmetric configuration and relatively gentle variation of frequency in single PCNC. (b) The frequency splittings under different coupling strengths in PT symmetric sensors.

# Conclusions

The PT symmetry breaking phenomenon is investigated in coupled nanobeam cavities. The dependence of the transition threshold on the gain/loss contrast and coupling strength is considered and we find that larger coupling strength will lead to larger transition threshold, which is well consonant with the theoretical predictions. The realization of unidirectional light propagation and possibility of achieving single mode laser operation are discussed. Moreover, the enhanced sensitivity for single particle detection when the concept is exploited in sensors is also perceived.

# Acknowledgment

This work is partially supported by the National Natural Science Foundation of China (Nos. 91233208 and 60990322), the National High Technology Research and Development Program (863 Program) of China (No. 2012AA012201), the Program of Zhejiang Leading Team of Science and Technology Innovation.

# References

1. C. M. Bender and S. Boettcher, "Real spectra in non-Hermitian Hamiltonians having PT symmetry," Phys. Rev. Lett. 80, 5243–5246 (1998).

2. R. El-Ganainy, K. G. Makris, D. N. Christodoulides, and Z. H. Musslimani, "Theory of coupled optical PT-symmetric structures," Opt. Lett. 32(17), 2632–2634 (2007).

3. H. F. Jones, "Analytic results for a PT-symmetric optical structure," J. Phys. A 45, 135306 (2012).

4. S. Klaiman, U. Günther, and N. Moiseyev, "Visualization of branch points in PT-symmetric waveguides," Phys. Rev. Lett. 101(8), 080402 (2008).

5. A. Guo, G. J. Salamo, D. Duchesne, R. Morandotti, M. Volatier-Ravat, V. Aimez, G. A. Siviloglou, and D. N. Christodoulides, "Observation of $\mathcal{PT}$-symmetry breaking in complex optical potentials," Phys. Rev. Lett. 103, 093902 (2009).

6. C. E. Ruter, K. G. Makris, R. El-Ganainy, D. N. Christodoulides, M. Segev, and D. Kip, "Observation of parity - time symmetry in optics," Nature Phys. 6, 192–195 (2010).

7. K. G. Makris, R. El-Ganainy, D. N. Christodoulides, and Z. H. Musslimani, "Beam dynamics in PT symmetric optical lattices," Phys. Rev. Lett. 100, 103904 (2008).

8. Derek D. Scott and Yogesh N. Joglekar, "Degrees and signatures of broken $\mathcal{PT}$ symmetry in nonuniform lattices," Phys. Rev. A 83(5), 050102 (2011).

9. E. M. Graefe and H. F. Jones, "PT-symmetric sinusoidal optical lattices at the symmetry breaking threshold," Phys. Rev. A 84(1), 013818 (2011).

10. A. Regensburger, C. Bersch, M. A. Miri, G. Onishchukov, D. N. Christodoulides, and U. Peschel, "Parity-time synthetic photonic lattices," Nature 488(7410), 167–171 (2012).

11. H. Benisty, A. Degiron, A. Lupu, A. De Lustrac, S. Chénais, S. Forget, M. Besbes, G. Barbillon, A. Bruyant, S. Blaize, and G. Lérondel, "Implementation of PT symmetric devices using plasmonics: principle and applications," Opt. Express 19(19), 18004–18019 (2011).

12. S. Longhi, "PT-symmetric laser absorber," Phys. Rev. A 82, 031801(R) (2010).

13. Y. D. Chong, L. Ge, and A. D. Stone, "PT-symmetry breaking and laser-absorber modes in optical scattering systems," Phys. Rev. Lett. 106(9), 093902 (2011).

14. M. Liertzer, L. Ge, A. Cerjan, A. D. Stone, H. E. Türeci, and S. Rotter, "Pump-induced exceptional points in lasers," Phys. Rev. Lett. 108(17), 173901 (2012).

15. M. Brandstetter, M. Liertzer, C. Deutsch, P. Klang, J. Schöberl, H. E. Türeci, G. Strasser, K. Unterrainer, and S. Rotter, "Reversing the pump dependence of a laser at an exceptional point," Nat Commun 5, 4034 (2014).

16. A. E. Miroshnichenko, B. A. Malomed, and Yu. S. Kivshar, "Nonlinearly-PT-symmetric systems: spontaneous symmetry breaking and transmission resonances," Phys. Rev. A 84(1), 012123 (2011).

17. Y. Lumer, Y. Plotnik, M. C. Rechtsman, and M. Segev, "Nonlinearly induced $\mathcal{PT}$ transition in photonic systems," Phys. Rev. Lett. 111, 263901 (2013).

18. S.-B. Lee, J. Yang, S. Moon, S.-Y. Lee, J.-B. Shim, S. W. Kim, J.-H. Lee, and K. An, "Observation of an exceptional point in a chaotic optical microcavity," Phys. Rev. Lett. 103, 134101 (2009).

19. Y. Choi, S. Kang, S. Lim, W. Kim, J. R. Kim, J. H. Lee, and K. An, "Quasieigenstate coalescence in an atom-cavity quantum composite," Phys. Rev. Lett. 104(15), 153601 (2010).

20. Jan Wiersig, "Structure of whispering-gallery modes in optical microdisks perturbed by nanoparticles," Phys. Rev. A 84(6), 063828 (2011).

21. B. Peng, Ş. K. Özdemir, F. Lei, F. Monifi, M. Gianfreda, G. L. Long, S. Fan, F. Nori, C. M. Bender, and L. Yang, "Parity–time-symmetric whispering-gallery microcavities," Nat. Phys. 10(5), 394–398 (2014).

22. M. A. Miri, P. LiKamWa, and D. N. Christodoulides, "Large area single-mode parity-time-symmtric laser amplifiers," Opt. Lett. 37, 764–766 (2012).

23. H. Hodaei, M.-A. Miri, M. Heinrich, D. N. Christodoulides, and M. Khajavikhan, "Parity-time-symmetric microring lasers," Science 346, 975–978 (2014).

24. L. Feng, Y. L. Xu, W. S. Fegadolli, M. H. Lu, J. E. Oliveira, V. R. Almeida, Y. F. Chen, and A. Scherer, "Experimental demonstration of a unidirectional reflectionless parity-time metamaterial at optical frequencies," Nat. Mater. 12(2), 108–113 (2012).

25. X. Yin and X. Zhang, "Unidirectional light propagation at exceptional points," Nat. Mater. 12, 175–177 (2013).

26. L. Chang, X. Jiang, S. Hua, C. Yang, J. Wen, L. Jiang, G. Li, G. Wang, and M. Xiao, "Parity–time symmetry and variable optical isolation in active-passive-coupled microresonators," Nat. Photonics 8, 524–529 (2014).

27. Jan Wiersig, "Enhancing the sensitivity of frequency and energy splitting detection by using exceptional points: application to microcavity sensors for single particle detection," Phys. Rev. Lett. 112(20), 203901 (2014).

28. Q. Quan, P. B. Deotare, and M. Loncar, "Photonic crystal nanobeam cavity strongly coupled to the feeding waveguide," Appl. Phys. Lett. 96, 203102 (2010).

29. Q. Quan and M. Loncar, "Deterministic design of high Q, small mode volume photonic crystal nanobeam cavities," Opt. Express 19, 18529–18542 (2011).

30. K. Foubert, B. Cluzel, L. Lalouat, E. Picard, D. Peyrade, F. de Fornel, and E. Hadji, "Influence of dimensional fluctuations on the optical coupling between nanobeam twin cavities," Phys. Rev. B 85(23), 235454 (2012).

31. P. B. Deotare, M. W. McCutcheon, I. W. Frank, M. Khan, and M. Loncar, "Coupled photonic crystal nanobeam Cavities," Appl. Phys. Lett. 95(3), 031102 (2009).

32. C. M. Bender, M. Gianfreda, Ş. K. Özdemir, B. Peng, and L. Yang, "Twofold transition in PT-symmetric coupled oscillators," Phys. Rev. A. 88, 1–8 (2013).

33. A. Hassan, H. Hodaei, W. Hayenga, M. Khajavikhan, and D. Christodoulides, "Enhanced Sensitivity in Parity-Time-Symmetric Microcavity Sensors," in Advanced Photonics 2015, OSA Technical Digest (online) (Optical Society of America, 2015), paper SeT4C.3.